\documentstyle[11pt]{article}    
\def\bc{\begin{center}}              \def\ec{\end{center}}
\def\beq{\begin{equation}}           \def\eeq{\end{equation}}
\def\bear{\begin{eqnarray}}         \def\eear{\end{eqnarray}}
\def\bt{\begin{tabular}}             \def\et{\end{tabular}}
\def\la{\langle}    \def\ra{\rangle}  \def\nn{\nonumber}
\def\dg{\dagger}    \def\ci{\cite}   \def\lb{\label}  \def\ld{\ldots}
        \def\pr{\prime}  
    \def\vs{\vspace}  
   
             \def\longrar{\longrightarrow}
                  \def\noi{\noindent}

\def\alf{\alpha}     
     
\def\Dlt{\Delta}     \def\dlt{\delta}
\def\lam{\lambda}    \def\Lam{\Lambda}
       \def\sig{\sigma}

\begin{document}

\title{ Generalizations of Heisenberg uncertainty relation}
\author{D.A. Trifonov\\
 Institute for nuclear research, \\
 72 Tzarigradsko chaussee, 1784, Sofia}

\date{}
\maketitle

quant-ph/0112028  

v.2: Mar 6, 2002

\abstract{
A survey on the generalizations of Heisenberg uncertainty relation and
a general scheme for their entangled extensions to several states and
observables is presented. The scheme is illustrated on the examples of
one and two states and canonical quantum observables, and spin and
quasi-spin components. Several new uncertainty relations are displayed.
PACS\, 0365H, 4250D, 0220  }

\section{Introduction} 

The uncertainty (indeterminacy) principle exhibits a fundamental manner in
which the quantum description of nature departs from the classical one.
It was introduced in 1927 by  Heisenberg \ci{H} who demonstrated the
impossibility of simultaneous precise measurement of the canonical quantum
observables $x$ and $p_x$ (the coordinate and the momentum) by positing an
approximate relation $\dlt p_x\dlt x \sim \hbar$.  Soon after the
Heisenberg paper appeared  Kennard proved \ci{KW} the
inequality (furthermore we use the dimensionless $p$ and $q$, $[p,q]=-i$)
\beq\lb{HWK} 
(\Dlt p)^2 (\Dlt q)^2 \,\geq\, 1/4,
\eeq
where $(\Dlt p)^2$ and $(\Dlt q)^2$ are the variances (dispersions) of
$p$ and $q$, defined for any observable $X$ by $(\Dlt
X)^2:= \la(X-\la X\ra)^2\ra$.  The inequality (\ref{HWK})
became known as the {\it Heisenberg uncertainty relation} (Heisenberg UR)
for the two canonical observables.

Generalization of inequality (\ref{HWK}) to the case of arbitrary two
observables (Hermitian operators $X$ and $Y$) was made by H. Robertson in
1929 \ci{R1},
\beq\lb{RUR1}
(\Dlt X)^2(\Dlt Y)^2 \,\,\geq\,\,
\frac{1}{4}\left|\la [X,Y]\ra\right|^2.
\eeq
Robertson inequality (\ref{RUR1}) became known again as the Heisenberg
UR. However we prefer, in view of the Robertson
contribution, to call it Heisenberg--Robertson (H--R) UR.
This inequality (or its particular case (\ref{HWK}))
became an irrevocable part of every textbook on quantum mechanics and it is
regarded as a rigorous formulation of the Heisenberg indeterminacy
principle.

The aim of the present paper is to consider symmetry properties of the
H--R UR (section 2) and its natural invariant generalizations to several
observables and several states (sections 2 and 3).  The first
generalization is that made by Schr\"odinger \ci{S1}.  In section 2 we
briefly analyze the invariance properties of H--R UR and Schr\"odinger UR.
Three basic UR's for $n$ observables and $2$ state are displayed in
subsection 3.2. In section 3.3  a general scheme for construction of UR's
for $n$ observables and $m$ states is provided.  The relation of the
conventional inequalities to the sets of the widely used canonical
coherent states \ci{GKS} and squeezed states \ci{obzori,obzor2} is also
reminded.

\section{Invariant generalizations of Heisenberg--Robertson relation}

The conventional UR's (\ref{HWK}) and (\ref{RUR1}) suffers from two
deficiencies.
The first one is that they are not form-invariant under
the linear transformations of operators. In particular (\ref{HWK}) is not
invariant under linear canonical transformations: if
(\ref{HWK}) is minimized in a state $|\psi\ra$, then the canonical
transformation (rotation on angle $\theta$ in phase space)
\beq\lb{LCT}
p^\pr = p\cos\theta + q\sin\theta,\quad
q^\pr = -p\sin\theta + q\cos\theta,
\eeq
violates the equality in (\ref{HWK}). So it makes
sense to look for other uncertainty inequalities, which are invariant
under rotation (\ref{LCT}). At the "level" of two second moments of $p$
and $q$ such inequality is
\beq\lb{newin1}
(\Dlt p)^2 + (\Dlt q)^2 \,\geq\, 1.
\eeq
This inequality is less precise than (\ref{HWK}): the minimization of
(\ref{newin1}) entails
the equality in (\ref{HWK}), the inverse being untrue. For two arbitrary
observables (\ref{newin1}) takes the form
\beq\lb{newin1a}
(\Dlt X)^2 + (\Dlt Y)^2 \,\geq\,|\la[X,Y]\ra|.
\eeq
If $X$ and $Y$ are elements of a Lie algebra $L$, then
their "linear canonical transformations" are the automorphisms in $L$.

The second point is that for {\it two} observables there are {\it three}
second order statistical moments --
the variances of each observable and their {\it covariance}, while only
the first two ones are taken into account in (\ref{RUR1}).  This fact was
first noted by Schr\"odinger \ci{S1}, who derived (using Schwartz
inequality) the more general inequality
\beq\lb{SUR}       
(\Dlt X)^2(\Dlt Y)^2 - (\Dlt XY)^2 \,\,\geq\,\,
\frac{1}{4}\left|\la [X,Y]\ra\right|^2,
\eeq
where $\Dlt XY$ denotes the covariance of $X$ and $Y$, $\Dlt XY = \la
XY+YX\ra/2 - \la X\ra\la Y\ra$.
In the classical probability theory the vanishing covariance
is a necessary (but not sufficient) condition for the statistical
independence of two random quantities.

In the case of coordinate and momentum observables the relation (\ref{SUR})
takes the shorter  form of
\beq\lb{SURpq}
(\Dlt q)^2(\Dlt p)^2 - (\Dlt qp)^2 \,\,\geq\,\, 1/4.
\eeq
Schr\"odinger inequality (\ref{SUR}) is more general and more precise
than that of Heisenberg--Robertson, eq. (\ref{RUR1}): the former is
reduced to the latter in states with vanishing covariance of $X$
and $Y$, and the equality in (\ref{RUR1}) entails the equality in
(\ref{SUR}), the inverse being untrue.

One can easily check that Schr\"odinger UR is invariant
under all linear canonical transformations of $p$ and $q$, including the
scale transformations.
From the three inequalities (\ref{HWK}), (\ref{newin1}) and (\ref{SURpq})
it is the Schr\"odinger one that is most precise and the most symmetric.
The inequality (\ref{newin1}) is the most unprecise one: the equality in
it entails the equality in both (\ref{SURpq}) and (\ref{HWK}).

The interest in Schr\"odinger relation has grown in the last two decades
in connection with the description and experimental realization of the
canonical {\it squeezed states} of the electromagnetic radiation
\ci{obzori,obzor2}. This family can be defined \ci{T93} as the unique set
of states that minimize inequality (\ref{SURpq}).
It was only recently realized \ci{obzor2}, that the famous canonical
{\it coherent states} (introduced in \ci{GKS}) can be uniquely defined as
states that minimize (\ref{newin1}).

\section{Generalizations to several observables and several states}

The uncertainty relations (\ref{HWK}), (\ref{RUR1}) and (\ref{SUR})
provide a quantitative limitations to the accuracy of measurement of two
incompatible observables in one and the same state. Two natural questions
related to inequalities (\ref{HWK}) -- (\ref{SURpq}), can be
immediately formulated:

(a) are there nontrivial generalizations to the
    case of several observables and one state?

(b) are there nontrivial generalizations to the case
    of one or several observables and several states?

By "nontrivial generalizations" I mean uncertainty inequalities, which can
not be represented as sums or products of those for two observables and
one state. Such UR's could be called {\it observable-} or {\it state-
entangled}.

\subsection{ Robertson inequalities for $n$ observables}

The positive answer to the first of the above two questions was given by
H. Robertson in 1934 \ci{R2}, who generalized (\ref{RUR1}) and (\ref{SUR})
to the case of $n$ observables.
For $n$ Hermitian operators $X_i$,
$i=1,\ldots,n$, Robertson established the inequality
($\vec{X}= (X_1,\ldots,X_n)$)
\beq\lb{RUR2}
\det\sig(\vec{X}) \geq \det C(\vec{X}),
\eeq
where $\sig$ is the uncertainty matrix, and $C$ is a matrix of mean
commutators of $X_i$, and $X_j$:\,\,
$$ \sig_{ij} = \mbox{$\frac 12$}\la X_iX_j+X_jX_i\ra - \la X_i\ra\la
X_j\ra,\,\, C_{ij} = -\mbox{$\frac i2$}\la [X_i,X_j]\ra.$$
The diagonal element $\sig_{ii}$ is just the variance of $X_i$, while
$\sig_{ij}$ is the covariance of $X_i$ and $X_j$. At $n=2$ inequality
(\ref{RUR2}) recovers (\ref{SUR}).

Robertson UR (\ref{RUR2}) is {\it form-invariant} with respect to
any nondegenerate linear real transformation of $n$ operators $X_i$, the
equality being {\it invariant}. Indeed, let $X^\pr_i = \lam_{ij}X_j$,
where $\Lam = (\lam_{ij})$ is non-degenerate. Then the uncertainty matrix
$\sig^\pr$ for $X^\pr_i$ in the same state is obtained as
\beq\lb{sig'}
\sig^\pr := \sig(\vec{X}^\pr) = \Lam\sig\Lam^T,
\eeq
where $\Lam^T$ is transposed $\Lam$. Similarly $C^\pr =
\Lam C\Lam^T$. Then we see that the equality $\det\sig^\pr = \det C^\pr$
follows from  $\det\sig = \det C$ and vice versa. Therefore (\ref{RUR2})
generalizes the full symmetry properties of Schr\"odinger relation
(\ref{SUR}) to the case of $n$ observables. Robertson also noted the
generalization of the less precise inequality (\ref{RUR1}):
from the Hadamard inequality and (\ref{RUR2}) he
immediately derived the inequality (to be called Hadamard--Robertson UR)

\beq\lb{RUR3}
(\Dlt X_1)^2\ldots (\Dlt X_n)^2 \geq \det C(\vec{X}).
\eeq
Here I provide the generalization of the most unprecise inequality
(\ref{newin1a}) to the case of $n$ arbitrary observables $X_i$:
\beq\lb{newin2a}
{\rm Tr}\,\sig(\vec{X}) = \mbox{$\sum_i^n$} (\Dlt X_i)^2 \,\geq\,
\frac{1}{n-1}\mbox{$\sum_{j>i}^n$} |\la [X_i,X_j]\ra|.
\eeq
This inequality holds for any $n$. For even $n$, $n=2m$ it can be enhanced,
\beq\lb{newin2b}
{\rm Tr}\,\sig(\vec{X}) \geq \mbox{$\sum_{\mu=1}^{m}$} |\la
[X_\mu,X_{m+\mu}]\ra|,
\eeq

Note that (\ref{RUR2}) is most precise and most symmetric: it is
form-invariant under any nondegenerate linear transformation
$\vec{X}\longrar \vec{X}^\pr = \Lam\vec{X}$. 
One can check that the most unprecise inequality (\ref{newin2a}) is invariant
under orthogonal linear transformations of $X_i$.
The intermediate precision inequality (\ref{RUR3}) 
is most unsymmetric: it is
invariant under special scale transformations $X_k\rightarrow X_k/\alf_k$,
$X_{m+k}\rightarrow \alf_k X_{m+k}$ only ($k=1,\ldots m$, $m \leq [n/2]$).

The problem of minimization of Robertson relation (\ref{RUR2}) is
considered in \ci{T97} (see also \ci{obzor2}). It is worth noting the
result that the group-related coherent states with maximal
symmetry (see \ci{KS} and references therein), the simplest examples of
which are the spin and the quasi-spin coherent states, are the unique
states that minimize Robertson inequality (\ref{RUR2}) for the Hermitian
components of the Weyl lowering and raising operators. Therefore these
states can be alternatively defined as Robertson minimum uncertainty
states  (called also Robertson intelligent states \ci{T97}) for several
observables.

On the example of $2m$ canonical observables $X_i=Q_i$, $Q_\mu=p_\mu$,
$Q_{m+\mu} = q_\mu$,
inequalities (\ref{RUR2}) and (\ref{RUR3}) and (\ref{newin2b}) read
\beq\lb{RUR2pq}
\det\sig(\vec{Q}) \geq  \frac{1}{4^m},
\eeq
\vs{-2mm}\beq\lb{RUR3pq}
(\Dlt p_1)^2(\Dlt q_1)^2\cdots(\Dlt p_m)^2(\Dlt q_m)^2
\geq \frac{1}{4^m},
\eeq
\vs{-2mm}\beq\lb{newin2pq}
{\rm Tr}\,\sig(\vec{Q}) =
\mbox{$\sum_{\mu=1}^m$} \left[(\Dlt p_\mu)^2 + (\Dlt q_\mu)^2\right]
 \geq  m.
\eeq
The equality in (\ref{RUR2pq}) is reached in the multimode squeezed states
\ci{obzori,obzor2}  (for this and other examples of states that minimize
(\ref{RUR2}) with $n>2$ see refs. \ci{T97,obzor2}).  The squeezed states with
vanishing covariances of all $p_\mu$ and $q_\nu$ minimize (\ref{RUR3}),
while (\ref{newin2pq}) is minimized only in multimode canonical coherent
states (in which one also has $\Dlt p_\mu q_\nu =0$).

Let us note that ${\rm Tr}\,\sig(\vec{Q})$ (and ${\rm
Tr}\,(\sig(\vec{Q}))^k$ as well) is invariant under orthogonal
transformations $\vec{Q}\longrar \Lam\vec{Q}$, but not under symplectic
ones.
Symplectic transformations preserve the canonical commutation relations.
In the one mode case, i.e.  $m=1$, all orthogonal matrices are symplectic.
Invariant under symplectic transformations is the quantity ${\rm
Tr}\,(\sig J)^k$.  In ref. \ci{SCB,T97} the following invariant
inequalities were proved (please note that in ref. \ci{SCB} factor $i$ is
omitted)
\beq\lb{SCB UR}
{\rm Tr}\,[i\sig(\vec{Q}) J]^{2k} \geq m/2^{2k-1}.
\eeq
At $m=1$ and $k=1$ Schr\"odinger UR (\ref{SURpq}) is recovered.

The $2m$ canonical operators $Q_i$ are known to close the Heisenberg--Weyl
Lie algebra. In the general case of $X_i$ closing any Lie algebra the
right hand sides of the above UR's (\ref{RUR2}), (\ref{RUR3}) --
(\ref{newin2b}) are, due to the commutation
relations $[X_i,X_k] = c_{ik}^j X_j$,  combinations of mean values of
$X_i$.
Let us note the case of $su(2)\sim so(3)$ and $su(1,1)\sim so(2,1)$
algebras.  The three basic operators $X_i=J_i$ of $su(2)$  are the spin
operators, and the three generators $K_i$ of 
$su(1,1)$ are known as quasi-spin operators: 
$$[J_k,J_l]=i\epsilon_{klm}J_m,\quad
[K_1,K_2]=-iK_3,\,  [K_2,K_3]=iK_1,\, [K_3,K_1]=-iK_2.$$ 
The inequality (\ref{newin2a}), applied to $J_i$ and $K_i$, tells us that
the sum of fluctuations (variances) of spin or quasi-spin components is
not less than the sum of their absolute mean values,  
$$ \sum_{i=1}(\Dlt J_i)^2 \geq \frac 12
\sum_{i=1}\left|\la J_i\ra\right|,
\quad  \sum_{i=1}(\Dlt K_i)^2 \geq \frac 12 \sum_{i=1}\left|\la
K_i\ra\right|.$$
\vs{2mm}

\subsection{Extension of Heisenberg and Schr\"odinger relations to two
distinct states}

All the uncertainty relations so far considered in the literature (and the
above (\ref{RUR2})--(\ref{SCB UR}) as well) have the form of
inequalities between first and second (or higher) moments of the
observables {\it in one and the same state}. Furthermore such uncertainty
relations should be called {\it conventional}.

However it is clear that one can measure  and compare the accuracy of
measurement of observables in distinct states. Therefore it is reasonable
to look for nontrivial inequalities between statistical moments of
observables in {\it two and several states}.
Such relation should be called {\it state-extended} or {\it
state-entangled}. On the lowest level of two observables such extended
inequalities can be obtained by somewhat elementary manipulations of the
Heisenberg--Robertson and/or Schr\"odinger inequalities written for two
distinct states. Here are the state-entangled extensions of the
conventional relations (\ref{newin1}), (\ref{HWK}) and (\ref{SUR}) for $p$
and $q$ to the case of two state $|\phi\ra$ and $|\psi\ra$,

\beq\lb{enewin1}
\Dlt_\psi q\,\Dlt_\phi p + \Dlt_\phi q\,\Dlt_\psi p \geq 1\, ,
\eeq
\beq\lb{ehur}
(\Dlt_\psi q)^2(\Dlt_\phi p)^2 + (\Dlt_\phi q)^2
(\Dlt_\psi p)^2 \geq\,\,\frac 12\, ,
\eeq
\beq\lb{esur}
(\Dlt_\psi q)^2(\Dlt_\phi p)^2 + (\Dlt_\phi q)^2
(\Dlt_\psi p)^2 - 2\left|\Dlt_\psi qp\,\Dlt_\phi qp\right|\,\,
\geq\,\, \frac 12\, ,
\eeq

\noi where $\Dlt_\psi XY$ is the covariance of $X$ and $Y$ in the state
$|\psi\ra$. At $|\psi\ra = |\phi\ra$ the conventional inequalities
(\ref{newin1}), (\ref{HWK}) and (\ref{SUR}) are recovered. Note the symmetry
under transpositions $p\leftrightarrow q$ and $|\psi\ra \leftrightarrow
|\phi\ra$.

\subsection{General scheme for uncertainty relations}

A quite general scheme for construction of uncertainty relations for $n$
observables and $m$ states is provided by the following Lemma,
\vs{1mm}

{\normalsize\bf Lemma.}
{\sl If $H_\mu$, $\mu=1,\ldots,m$ are non-negative definite
Hermitian $n\times n$ matrices, then the following inequalities hold
\bear
{\cal M}\left(i_1,\ld,i_r;\mbox{$\sum_\mu S_\mu$}\right) &\,\geq\,& {\cal
M}\left(i_1,\ld,i_r; \mbox{$\sum_\mu A_\mu$}\right), \quad \lb{M_r 1}\\
{\cal M}\left(i_1,\ld,i_r;\mbox{$\sum_\mu H_\mu$}\right) &\,\geq\,& \sum_\mu{\cal
M}(i_1,\ld,i_r;H_\mu), \lb{M_r 2}
\eear
\vspace{-5mm}\bear
{\rm Tr}\,S_\mu &\,\geq\,& \frac{2}{n-1}\,
\sum_{j>i}^n\left|(A_\mu)_{ij}\right|\quad \mbox{ for any $n$},
\lb{M_r 3} \\
{\rm Tr}\,S_\mu &\,\geq\,& \sum_{\nu=1}^{m}\left|(A_\mu)_{\nu,m+\nu}\right|
\quad \mbox{for even $n=2m$},\lb{M_r 4}
\eear
where ${\cal M}\left(i_1,\ld,i_r;B\right)$ is the principal minor of order
$r\leq n$ of matrix $B$, and $S_\mu$ and $A_\mu$ are the symmetric and the
antisymmetric part of $H_\mu$: $H_\mu = S_\mu + iA_\mu$.}
\vs{1mm}

The first two inequalities (\ref{M_r 1}) and (\ref{M_r 2}) are proved in
the second paper of ref. \ci{T00} (Lemma 2 in \ci{T00}).\,
The proof of the third matrix inequality of the Lemma can be performed
along the following lines. We
represent the trace of $S$ in the form
$ {\rm Tr}\,S = (1/(n-1))\sum_{j>i}(S_{ii}+S_{jj})$
and consider the $2\times2$ principal submatrices $S(ij)$ of $S$,
$[S(ij)]_{11} = S_{ii}$, $[S(ij)]_{12} = S_{ij} = [S(i,j)]_{21}$,
$[S(jj)]_{22} = S_{jj}.$
These are symmetric and non-negative definite \ci{T00,Gant}.
Similarly we define $2\times2$ antisymmetric
matrices $A(ij)$ and compose $H(ij) = S(ij) +iA(ij)$. Then we apply
(\ref{M_r 1}) to $H(ij)$ and use
the Hadamard inequality to  obtain
$S_{ii}+S_{jj} \geq 2\left|A_{ij}\right|$,
wherefrom the desired inequality (\ref{M_r 3}) follows.\,\,

For even $n$, $n=2m$, we put ${\rm Tr}S = \sum_{\nu=1}^{m}(S_{\nu\nu} +
S_{\nu+m,\nu+m})$ and in a similar way obtain inequality (\ref{M_r 4}).

For brevity we shall call the above inequalities {\it principal
minor inequalities} of type $(n,m)$ and order $r$.

Since the characteristic coefficients $C_r^{(n)}(B)$ of order $r\leq n$ of
any matrix $B$ are sums of the principal minors \ci{Gant}, the above
inequalities (\ref{M_r 1}) and (\ref{M_r 2}) remain valid if one replaces
${\cal M}\left(i_1,\ld,i_r;B\right)$ with $C_r^{(n)}(B)$ for the
corresponding $B$.
(For example, in the left hand side of the first
inequality (\ref{M_r 1}) $B=\sum_\mu S_\mu$, and in the right hand side
$B=\sum_\mu A_\mu$). 
The obtained inequalities for $C_r^{(n)}(B)$ are called {\it characteristic}.
Note that ${\cal M}\left(i_1,\ld,i_n;B\right)= C_n^{(n)}(B)$ and 
$C_n^{(n)}(B)= \det B$, while $C_n^{(1)}(B) = {\rm Tr}\,B$.

By a suitable physical choice of matrices $H_\mu$ in matrix inequalities
(\ref{M_r 1}) -- (\ref{M_r 4}), and in the corresponding characteristic
inequalities as well,  one can obtain series of physical relations.  If
matrices $H_\mu$ are constructed by means of statistical moments of
observables, then the obtained inequalities are UR's, which could
naturally be called  {\it principal} or {\it characteristic} UR's.

All UR's considered in the previous sections and subsections, including
(\ref{SCB UR}), can be casted in the forms (\ref{M_r 1}) -- (\ref{M_r 4}).
The conventional UR (\ref{SCB UR}) is of the form (\ref{M_r 4}) with $S =
(i\sig^\pr(\vec{Q}) J)^{2k}$ and $A=(A_{ij})$:
$A_{\mu\nu}=A_{m+\mu,m+\nu} = 0$, $A_{\nu,m+\nu} = - A_{m+\nu,\nu} =
1/2^{2k-1}$, where $\sig^\pr$ is the diagonal uncertainty matrix, obtained
from $\sig$ by means of a symplectic transformation $\vec{Q}^\pr =
\Lam\vec{Q}$ \ci{T97,SCB}.
Next we consider some illustrative examples of non-negative
matrices $H_\mu=H^\dg_\mu$, $\mu=1,\ldots,m$, and the related new UR's.
\vs{2mm}

{\normalsize\bf Example 1.} An illustrative physical choice of $H_\mu$ is
the following $H_\mu = G_\mu(\vec{X};\psi_\mu)$, where
\beq\lb{choice 1}
G_{\mu;ij} = \la (X-\la X_i\ra)\psi_\mu|
(X_j-\la X_j\ra)|\psi_\mu\ra.
\eeq
Matrix $G_\mu$ can be recognized as Gram matrix for the transformed states
$|\chi_{\mu i}\ra = (X-\la X_i\ra)|\psi_\mu\ra$. Its symmetric part
is defined as the uncertainty matrix $\sig$. For one state and $n$
operators $X_i$ the inequality (\ref{M_r 1}) with
$r=n$ and $H = G(\vec{X};\psi)$ coincides with
Robertson inequality (\ref{RUR2}). For
two states and $2$ observables $p$
and $q$ the senior inequalities (\ref{M_r 1}) and (\ref{M_r 2})
coincide and (with $H_\mu = G(\vec{Q};\psi_\mu)$) reproduce (\ref{esur});
for arbitrary two $X$ and $Y$ (\ref{M_r 1}) produces

\bear\lb{(2,2)Sa1}         
(\Dlt_{\psi_1}X)^2(\Dlt_{\psi_2}Y)^2 + (\Dlt_{\psi_2}X)^2
(\Dlt_{\psi_1}Y)^2 \hspace{25mm}\nn \\ 
- 2\left|\Dlt_{\psi_1}XY\Dlt_{\psi_2}XY\right| 
\,\geq\,  \frac
12\left|\la[X,Y]\ra_{\psi_1}\la[X,Y]\ra_{\psi_2}\right|.\quad
\eear
This is a direct extension of conventional Schr\"odinger UR (\ref{SUR}) to
two distinct states.
\vs{1mm}

{\normalsize \bf Example 2.}
Another interesting family of uncertainty relations is provided by the choice
$H_\mu = \,^\pr G_\mu(\vec{X};\psi_\mu;k)$, where
\beq\lb{choice 2}
\,^\pr G_{\mu;ij} = \la
(X_i^k-\la X_i\ra^k)\psi_\mu|  (X_j^k-\la X_j\ra^k)|\psi_\mu\ra,
\eeq
where $X_j^k$ is the $k$-th power of $X_j$.
Matrix $\,^\pr G_\mu$ can be recognized as Gram matrix for the transformed
states $|\,^\pr\chi_{\mu,i}\ra = (X_i^k-\la X_i\ra^k)|\psi_\mu\ra$. The
diagonal elements of matrix $\,^\pr G_\mu$ are nothing but the
$k$-order statistical moments of $X_i$. Thus the resulting principal
inequalities are {\it uncertainty relations for higher statistical
moments} of $n$ observables in $m$ states.
\vs{2mm}

Gram matrix $G$ for any $n$ vectors $|\chi_\mu\ra = a_{\mu
k}(X;\psi_k)|\psi_k\ra$ ($a_{\mu k}(X,\psi_k)$ being combinations of
operators $X_i$ and their moments in $|\psi_k\ra$) 
will produce some UR \ci{T00}. Note that for one state $G =
\la\chi|\chi\ra$ with $\det G =  \la\chi|\chi\ra \geq 0$. 
For example,  take the combination \ci{Flem} $|\chi\ra = |\phi\ra - 
|\psi\ra\la\psi|\phi\ra - |\psi_X\ra\la\psi_X|\phi\ra$, where
$|\psi_X\ra = (X\!-\!\la X\ra_\psi)|\psi\ra/\Delta_\psi X$.
Now $\det G \geq 0$ immediately produces the inequality 
$$ \Delta_\psi X \geq \left|\la\phi|(X\!-\!\la X\ra_\psi)|\psi\ra\right|^2 
/\sqrt{1\!-\!\left|\la\phi|\psi\ra\right|^2} \equiv f[X,\psi,\phi].$$

Similarly $\Delta_{\phi}X = f[X,\phi,\psi]$ ($\psi$ and $\phi$
interganged) and 
$$\Delta_\psi X\Delta_{\phi}X \geq f[X,\psi,\phi]\,f[X,\phi,\psi] \geq 0.$$

\section{Concluding remarks}

The presented scheme for construction of uncertainty relations is quite
general, since it is based on abstract matrix inequalities (\ref{M_r 1}) --
(\ref{M_r 4}).
Any non-negative Hermitian matrices $H_\mu$ involving (second or higher)
statistical moments of observables, both for quantum and
classical stochastic observables, can be used in this scheme to produce a
hierarchy of uncertainty relations of types (n,m).  We have shown that the
basic quantity for the uncertainty relations in quantum physics is the
Gram matrix. Its symmetric part can be regarded as a generalization of the
uncertainty matrix $\sig(\vec{X})$. This is most clear for Gram matrix of
the form (\ref{choice 1}). This definition of $\sig(\vec{X})$ persists for
the states that are outside the domain of the product $X_iX_j$, and is
valid for mixed states as well.
\vs{5mm}

\end{document}